\documentclass[12pt]{article}
\usepackage{epsfig, amsmath, amssymb}
\usepackage{graphicx,psfrag,float,color} 
\newcommand{\be}{\begin{equation}}
\newcommand{\ee}{\end{equation}}
\newcommand{\bea}{\begin{eqnarray}}
\newcommand{\eea}{\end{eqnarray}}
\newcommand{\bA}{\begin{array}}
\newcommand{\eA}{\end{array}}
\newcommand{\bc}{\begin{center}}
\newcommand{\ec}{\end{center}}
\newcommand{\al}{\alpha}

\newcommand{\ra}{\rightarrow}
\newcommand{\del}{\partial}

\newcommand{\ie}{{\it i.e.}}
\newcommand{\eg}{{\it e.g.}}

\newcommand{\delB}{\Delta B}

\newcommand{\cO}{{\cal O}}
\newcommand{\lA}{\langle}
\newcommand{\rA}{\rangle}

\begin{document}


\begin{titlepage}

\bc

\hfill 
\\         [22mm]

{\Huge $AdS$ plane waves, entanglement\\ [2mm] and mutual information}
\vspace{16mm}

{\large Debangshu Mukherjee and K.~Narayan} \\
\vspace{3mm}
{\small \it Chennai Mathematical Institute, \\}
{\small \it SIPCOT IT Park, Siruseri 603103, India.\\}

\ec
\medskip
\vspace{40mm}

\begin{abstract}
$AdS$ plane wave backgrounds are dual to CFT excited states with
energy momentum density $T_{++}=Q$. Building on previous work on
entanglement entropy in these and nonconformal brane plane wave
backgrounds, we first describe a phenomenological scaling picture for
entanglement in terms of ``entangling partons''. We then study aspects
of holographic mutual information in these backgrounds for two strip
shaped subsystems, aligned parallel or orthogonal to the flux. We
focus on the wide ($Ql^d\gg 1$) and narrow ($Ql^d\ll 1$) strip
regimes. In the wide strip regime, mutual information exhibits growth
with the individual strip sizes and a disentangling transition as the
separation between the strips increases, whose behaviour is distinct
from the ground and thermal states. In the narrow strip case, our
calculations have parallels with ``entanglement thermodynamics'' for
these $AdS$ plane wave deformations. We also discuss some numerical
analysis.
\end{abstract}

\end{titlepage}

{\tiny 
\begin{tableofcontents}
\end{tableofcontents}
}


\section{Introduction}

Inspired by the area scaling of black hole entropy, Ryu and
Takayanagi \cite{Ryu:2006bv,Ryu:2006ef}, \cite{HEEreview} identified a
simple geometric prescription for entanglement entropy (EE) in field
theories with gravity duals: the EE for a subsystem in the $d$-dim 
field theory is the area in Planck units of a minimal surface bounding 
the subsystem, the bulk theory living in $d+1$-dimensions. 
This is a prescription in the large $N$ classical
gravity limit. In recent times, entanglement entropy has been explored
widely, the holographic prescription giving a calculable handle on
what in field theory is a rather complicated question. For non-static
situations, the prescription generalizes to finding the area of an
appropriate bulk extremal surface with minimal area \cite{HRT}.

We are interested in studying excited states of a certain kind in this
paper, building on previous work. $AdS$ plane waves
\cite{Narayan:2012hk} \cite{Singh:2012un} \cite{Narayan:2012wn} are
deformations of $AdS$ which are dual to CFT excited states with
constant energy-momentum flux $T_{++}\sim Q$ turned on. Upon
$x^+$-dimensional reduction, these give rise to hyperscaling violating
spacetimes \cite{HV}, some of which exhibit
violations \cite{Ogawa:2011bz,Huijse:2011ef,Dong:2012se} of the area
law \cite{AreaLaw}.  In \cite{Narayan:2012ks}, a systematic study of
entanglement entropy for strip subsystems was carried out in $AdS$
plane waves (with generalizations to nonconformal brane plane waves in
\cite{Narayan:2013qga}).  The EE depends on the orientation of the
subsystem \ie\ whether the strip is parallel or orthogonal to the flux
$T_{++}$.  For the strip subsystem along the flux, the EE grows
logarithmically with the subsystem width $l$ for the $AdS_5$ plane wave
(the corresponding hyperscaling violating spacetime lies in the family
giving log-behaviour). The $AdS_4$ plane wave dual to plane wave 
excited states in the M2-brane Chern-Simons CFT exhibits an even 
stronger $\sqrt{l}$ growth. For the strip orthogonal to the flux, we 
have a phase transition with the EE saturating for $l\gg Q^{-1/d}$.

For two disjoint subsystems, an interesting information-theoretic object 
is mutual information (MI), defined as
\be
I[A,B]=S[A]+S[B]-S[A\cup B]\ ,
\ee
involving a linear combination of entanglement entropies.
It measures how much two disjoint subsystems are correlated (both 
classical and quantum). The EE terms in $I[A,B]$ automatically cancel
out the cutoff-dependent divergence thus making MI finite and positive
semi-definite. A new divergence comes up when the subsystems
collide. The term $S[A\cup B]$ in the above expression depends on the
separation between the subsystems $A$ and $B$: in the holographic
context, there are two extremal surfaces of key interest.  For large
separation, the disconnected surface $S[A\cup B]=S[A]+S[B]$ having
lower area is the relevant surface so that mutual information $I[A,B]$
vanishes.  For nearby subsystems however, the connected surface has
lower area.  Thus the Ryu-Takayanagi prescription automatically
implies a disentangling transition for mutual information in this
large $N$ classical gravity approximation \cite{Headrick:2010zt}, with
a critical separation $x_c$.

In this paper, we first discuss a phenomenological scaling picture for
entanglement for CFT ground and some excited states, building on some
renormalization-group like intuition described in
\cite{Swingle:2010jz} based on ``entangling bits'' or ``partons''
(sec. 3). In sec. 4 we describe some generalities on holographic
mutual information and then study mutual information in $AdS$ plane
waves for two parallel disjoint strip subsystems of width $l$ each
(sec. 5), first discussing the wide strip regime $Ql^d\gg 1$,
exhibiting again a disentangling transition. Then we study the
perturbative regime $Ql^d\ll 1$ and calculate the changes in the
turning point and the entanglement area functional to $O(Q)$ treating
the $AdS$ plane wave as a perturbation to pure $AdS$, for the strip
subsystem both parallel and orthogonal to the energy-momentum
flux. This perturbative analysis has parallels with ``entanglement
thermodynamics'' \cite{Bhattacharya:2012mi}
\cite{Allahbakhshi:2013rda} \cite{Wong:2013gua}. Finally, we perform
some numerical analysis to gain some insights when $Ql^d$ is
$O(1)$. We discuss some similarities and key differences of our
investigations with the study of mutual information for thermal
excited states \cite{Fischler:2012uv}, which are somewhat different
from these pure excited $AdS$ plane wave states. Sec. 2 contains a
review of $AdS$ plane waves and entanglement entropy.

\section{Review: $AdS$ plane waves and entanglement entropy}

$AdS$ plane waves \cite{Narayan:2012hk} \cite{Singh:2012un} 
\cite{Narayan:2012wn} are rather
simple deformations of $AdS/CFT$, dual to anisotropic excited states
in the CFT with uniform constant energy-momentum density $T_{++}$
turned on (with all other energy-momentum components vanishing),
\be\label{adsplanewave}
ds^2 = {R^2\over r^2} (-2dx^+dx^-+dx_i^2+dr^2)
+ R^2 Q r^{d-2} (dx^+)^2 + R^2 d\Omega^2\ ,
\ee
with $d$ the boundary spacetime dimension and
$R^4\sim g_{YM}^2 N {\alpha'}^2$ [$AdS_5$ plane wave],\ 
$R^6\sim Nl_P^6$ [$AdS_4$ plane wave]. These are normalizable 
deformations of $AdS_{d+1}\times S$ that arise in the near horizon 
limits of various conformal branes in string/M-theory. 
Structurally they are similar to the $AdS$ null deformations 
\cite{Balasubramanian:2010uk,Donos:2010tu} that give rise to 
gauge/string realizations of $z=2$ Lifshitz spacetimes 
\cite{Kachru:2008yh,Taylor:2008tg}, except that these $AdS$ plane waves 
are normalizable null deformations.
Reducing on the sphere, these are solutions in a $d+1$-dim effective 
gravity theory with negative cosmological constant and no other matter, 
\ie\ satisfying\ $R_{MN}=-{d\over R^2} g_{MN}$.\ The parameter $Q > 0$ 
gives rise to a holographic energy-momentum density $T_{++}\propto Q$ 
in the boundary CFT.
Dimensionally reducing (\ref{adsplanewave}) on the $x^+$-dimension
(and relabeling $x^-\equiv t$) gives a hyperscaling violating metric\ 
$ds^2=  r^{2 \theta \over d_i} \big(-{dt^2 \over r^{2z}} + 
{\sum_{i=1}^2  dx_i^2 + dr^2 \over r^2 }\big),$\ with exponents\ 
$z={d-2\over 2}+2 ,\ \theta={d-2\over 2}$\ and $d_i$ is the boundary 
spatial dimension. These are conformal to Lifshitz space times and appear 
in various discussions of non-relativistic holography, arising in 
various effective Einstein-Maxwell-scalar theories \eg\ \cite{HV}: 
see \cite{Dong:2012se} for various aspects of holography with 
hyperscaling violation. It is known that these spacetimes for the
special family ``$\theta=d_i-1$'' exhibit a logarithmic violation of 
the area law \cite{AreaLaw} of entanglement entropy, suggesting that 
these are signatures of hidden Fermi surfaces 
\cite{Ogawa:2011bz,Huijse:2011ef}. For the special case 
of the $AdS_5$ plane wave, we have $\theta=1, d_i=2$, lying in this 
``$\theta=d_i-1$'' family.

This spacetime (\ref{adsplanewave}) can be obtained
\cite{Singh:2012un} as a ``zero temperature'', highly boosted, 
double-scaling limit of
boosted black branes, using \cite{Maldacena:2008wh}. For instance, 
$AdS_5$ Schwarzschild black brane spacetimes, with metric\
$ds^2={R^2\over r^2} [-(1-r_0^4r^4) dt^2 + dx_3^2 +\sum_{i=1}^2
dx_i^2] + R^2 {dr^2\over r^2 (1-r_0^4r^4)}$ can be recast in boundary 
lightcone coordinates $x^\pm$ with 
$t={x^++x^-\over\sqrt{2}},\ x_3={x^+-x^-\over\sqrt{2}}$. After 
boosting by $\lambda$ as\ $x^\pm\ra \lambda^{\pm 1}x^\pm$, we obtain
$ds^2 = {R^2\over r^2} \left[-2dx^+dx^-+ {r_0^4 r^4\over 2}
(\lambda dx^++\lambda^{-1} dx^-)^2 +\sum_{i=1}^2 dx_i^2\right]
+ R^2 {dr^2\over r^2 (1-r_0^4r^4)}$~.
Now in the double scaling limit  
$r_0\ra 0 , \ \ \lambda\ra\infty$,\ with\ $Q={r_0^4\lambda^2\over 2}$ 
fixed, this becomes (\ref{adsplanewave}).
For the near extremal $AdS$ plane wave, from \cite{Maldacena:2008wh}, 
we see that we have other energy-momentum components also turned on,\
$T_{++}\sim \lambda^2 r_0^4\sim Q ,\ \
T_{--}\sim {r_0^4\over\lambda^2}\sim {r_0^8\over Q}~,
\ T_{+-}\sim r_0^4 ,\ T_{ij}\sim r_0^4\delta_{ij}$.\ 
Turning on a small $r_0$ about (\ref{adsplanewave}), this means
$T_{++}$ is dominant while the other components are small. In some
sense, this is like a large left-moving chiral wave with $T_{++}\sim Q$,
with a small amount of right-moving stuff turned on. Thus the
near-extremal case (with small $r_0$) serves to regulate the $AdS$
plane wave in the deep interior.

We now review certain aspects of holographic entanglement entropy
in these $AdS$ plane wave geometries \cite{Narayan:2012ks}. First,
it is worth recalling that the entanglement entropy for ground states
($Q=0$) in the d-dim CFTs arising on the various conformal branes with
strip-shaped subsystems has the form (upto numerical coefficients)
\be\label{EEgndconf}
S_A\ \sim\ {R^{d-1}\over G_{d+1}} \Big( {V_{d-2}\over\epsilon^{d-2}} - c_d
{V_{d-2}\over l^{d-2}}\Big)\ ,\qquad {R^3\over G_5}\sim N^2\ (4d\ CFT),
\ \ {R^2\over G_4}\sim N^{3/2}\ (3d\  CFT)\ ,
\ee
where $c_d>0$ is some constant, $l$ the strip width, $V_{d-2}$ the
longitudinal size and $\epsilon$ the ultraviolet cutoff.\ (We have used 
the relations\
$R_{D3}^4\sim g_sNl_s^4,\ R_{M2}^6\sim Nl_P^6$, and those for the Newton
constants\ $G_{10}\sim G_5 R_{D3}^5, G_{11}\sim G_{4} R_{M2}^{7}$, where
$g_s$ is the string coupling, and $l_s, l_P$ the string and Planck
lengths.)\  The first term exhibiting the leading divergence
represents the area law while the second term is a finite
cutoff-independent part encoding a size-dependent measure of the
entanglement \cite{Ryu:2006bv,Ryu:2006ef,CH}. With $Q\neq 0$, we have
an energy flux in a certain
direction: these are nonstatic spacetimes, and we therefore use the
covariant formulation of holographic entanglement entropy \cite{HRT}
working in the higher dimensional theory (with $x^+$ noncompact),
the strip geometry corresponding to a space-like subsystem on the
boundary.
Consider the strip to be along the flux direction, \ie\ with width
along some $x_i$ direction \cite{Narayan:2012ks}. Then the leading
divergent term is the same as for ground states. The width scales
as\ $l\sim r_*$, where $r_*$ is the turning point of the bulk extremal
surface, and the finite cutoff-independent piece in these excited states is
\bea\label{EEfinAdspw}
&& \qquad\quad \pm
\sqrt{Q} V_{d-2} l^{2-{d\over 2}}\ {R^{d-1}\over G_{d+1}}\qquad\qquad
[+:\ d<4,\ \ \ -:\ d>4]\ ;\nonumber\\
&& \sqrt{Q}V_2 N^2\ \log (lQ^{1/4})\ \ (D3)\ ;\qquad
\sqrt{Q} L \sqrt{l}\ N^{3/2}\ \ (M2)\ .
\eea
Note that the logarithmic behavior for the 4-dim CFT is of the same
form as for a Fermi surface, if the energy scale $Q^{1/4}$ is identified
with the Fermi momentum $k_F$. Both 4- and 3-dim CFTs in these excited
states thus exhibit a finite entanglement which grows with subsystem
size $l$. In particular, for fixed cutoff, this finite part is larger
than the leading divergence.  Recalling that the finite entanglement
for the thermal state (\ie\ the $AdS$ black hole) is extensive, of the
form\ $V_{d-2} T^{d-1} l$, we see that these are states with
subthermal entanglement. These are pure states in the large $N$
gravity approximation since the entropy density vanishes.

It is worth noting that we regard the $AdS$ plane wave spacetimes as a
low temperature highly boosted limit of the $AdS$ black brane: the
scale $Q=\lambda^2 r_0^4 \gg r_0^4$ implies a large separation of
scales between the flux in the $AdS$ plane wave and the temperature of
the black brane, with $Q$ dominating the physics in the plane wave
regime. The above estimates (\ref{EEfinAdspw}) for the finite part of
entanglement arise if the bulk extremal surface dips deep enough in
the radial direction to experience substantial deviation from the
$AdS$ geometry due to the plane wave, while still away from the
regulating black brane horizon in the deep interior, \ie\ the length
scales satisfy\ $Q^{-1/d} \ll l \ll {1\over r_0}$. 

With the strip orthogonal to the flux direction, a phase
transition was noted \cite{Narayan:2012ks}: for large width $l$,
there is no connected surface corresponding to a space-like subsystem, 
only disconnected ones.

This analysis can be extended \cite{Narayan:2013qga} to the various
nonconformal Dp-brane systems \cite{Itzhaki:1998dd}. These have a
ground state entanglement \cite{Ryu:2006ef,Barbon:2008ut} (after
converting to field theory parameters)\
$S_A = N_{eff}(\epsilon) {V_{d-2}\over\epsilon^{d-2}} -
c_d N_{eff}(l) {V_{d-2}\over l^{d-2}}$,\ with a scale-dependent number
of degrees of freedom\
$N_{eff}(l)=N^2 \left({g_{YM}^2N\over l^{p-3}}\right)^{{p-3\over 5-p}}$\
involving the dimensionless gauge coupling at scale $l$.
For nonconformal Dp-brane plane waves, it turns out to be natural
to redefine the energy density as $Q\ra Q N_{eff}(l)$ (\ie\ $Q$ in the
conformal cases above is the energy density per nonabelian degree of
freedom), and then the finite part of entanglement takes the form\
$S_A^{finite}\ \sim\ {\sqrt{N_{eff}(l)}\over 3-p}\
{V_{p-1}\sqrt{Q} \over l^{(p-3)/2}}$~ involving a dimensionless ratio
of the energy density and the strip width/lengths and $N_{eff}(l)$ \
(the leading divergence is as for the ground state). This finite part
is similar in structure to that for the conformal plane waves above,
but is scale-dependent: analysing the UV-IR Dp-brane phase diagram
\cite{Itzhaki:1998dd} shows the finite part to be consistent with
renormalization group flow \cite{Narayan:2013qga}.

\section{A phenomenological scaling picture for entanglement}

This is a generalization of an RG-like scaling picture in
\cite{Swingle:2010jz} for ground states. We assume a renormalization
group type scaling behaviour with a notion of ``entanglement per
scale'' as an organizing principle: \ie\ in a CFT of spacetime
dimension $d$, there are ``entangling bits'' or ``partons'' of all
sizes $s$. Equivalently at scale $s$, we think of space as
lattice-like with cell size $s$. In the ground state, each cell
roughly contains one entangling parton. Entanglement arises from
degrees of freedom straddling the boundary between the subsystem and
the environment, in other words from partons partially within the
subsystem and partly outside.  Entanglement entropy arises from the
fact that we trace over the environment and thus lose some information
about the straddling partons. The scaling picture below is admittedly
quite phenomenological and is only meant as an attempt at an intuitive
picture that fits the holographic entanglement calculations.

We want to estimate the rough number of degrees of freedom
contributing to entanglement at the interface between the subsystem
and the environment which has area $V_{d-2}\equiv L^{d-2}$.  At scale
$s$, the rough number of cells of linear size $s$ at the boundary is\
$({L\over s})^{d-2}={V_{d-2}\over s^{d-2}}$~. For a CFT with nonabelian
$N\times N$ matrix degrees of freedom, there are $N^2$ degrees of
freedom per cell (we use $N^2$ with a SYM CFT in mind but this can 
be easily generalized to $N^{3/2}$ for the M2-brane CFT). 
We then integrate this over all scales greater than the 
UV cutoff $\epsilon$ with the logarithmic measure ${ds\over s}$ 
and also we expect the IR cutoff is set by the subsystem size $l$. 
This gives (assuming $d>2$)
\be
S \sim \int_\epsilon^l {ds\over s}\ {V_{d-2}\over s^{d-2}} N^2\ 
\sim {N^2 V_{d-2}\over d-2} 
\left({1\over \epsilon^{d-2}} - {1\over l^{d-2}}\right) .
\ee
This shows the leading area law divergence and the 
subleading cutoff-independent finite part. For $d=2$, we obtain\ 
$S\sim \int_\epsilon^l {ds\over s}\ N^2\sim\ N^2 \log {l\over\epsilon}$ 
which is the logarithmic behaviour characteristic of a 2-dim CFT: this 
can be used as a check that the logarithmic measure ${ds\over s}$ is 
appropriate. This is a quantum entanglement, with contributions from 
various scales $s$.

Thus we see that there is a diverging number ${V_{d-2}\over s^{s-2}}$
of ultra-small partons at short distances $s\ra 0$ which essentially
gives rise to the area law divergence \cite{AreaLaw}. For excited
states, the energy-momentum density does not change the short distance
behaviour but implies an enhanced number of partons at length scales
much larger than the scale set by the energy-momentum, changing the IR
behaviour of entanglement as we will see below.

Similar arguments can be made for the various nonconformal gauge theories 
arising on the various nonconformal Dp-branes. Now the gauge coupling 
is dimensionful and the number of nonabelian degrees of freedom at 
scale $s$ is 
\be
N_{eff}(s) = N^2 \left({g_{YM}^2N\over s^{p-3}}\right)^{{p-3\over 5-p}}\ .
\ee
For the ground state, the entanglement at the boundary of 
the subsystem is obtained as before by integrating over all scales 
the number ${\cal N}_{eff}(s)$ of entangling bits or partons at scale $s$ 
\be
S \sim \int_\epsilon^l {ds\over s} {V_{d-2}\over s^{d-2}}  N_{eff}(s)\ 
\ \sim\ \ (5-p) N_{eff}(\epsilon) {V_{d-2}\over\epsilon^{d-2}}
\ -\ (5-p) N_{eff}(l) {V_{d-2}\over l^{d-2}}\ ,
\ee
in agreement with the known holographic result for the ground state 
entanglement for the nonconformal brane theories, upto numerical factors.
We see that the entanglement expression above breaks down for $p=5$: 
these are nonlocal theories (\eg\ little string theories for NS5-branes).

For the CFT$_d$ at finite temperature $T$, the entanglement entropy 
has a finite cutoff-independent piece which is extensive and dominant 
in the IR limit of large strip width $l$: this is the thermal entropy, 
essentially a classical observable,
\be
S \sim\ N^2 V T^{d-1} = N^2 {V\over (1/T)^{d-1}}\ ,\quad {\rm and}\quad
\rho \equiv {E\over V} \sim N^2 T^d\ ,
\ee
with $\rho$ the energy density and we have used 
${1\over T}= {\del S\over\del E}$~.
The energy density per nonabelian particle is\ 
${\rho\over N^2}=T^d={T\over (1/T)^{d-1}}$~, which suggests that the 
characteristic size of the typical particle is ${1\over T}$ with energy 
$T$. The CFT 
physics below this length scale ${1\over T}$~, in particular that of 
entanglement, will be indistinguishable from the 
ground state. Above this length scale, the presence of the energy 
density implies a larger number of entangling bits or partons and 
so a correspondingly larger entanglement. Thus the number of entangling 
partons ${\cal N}(s)$ for cell sizes $s\gg {1\over T}$ is the number 
of partons of individual volume $(1/T)^{d-1}$ in the total cell volume 
$s^{d-1}$, \ie\ ${\cal N}(s)|_{s\gg T^{-1}} \sim N^2 {s^{d-1}\over (1/T)^{d-1}}$:
thus ${\cal N}(s)$ is extensive for length scales larger than the 
inverse temperature. This implies a total entanglement 
\bea
S &\sim& \int_\epsilon^l {ds\over s} {V_{d-2}\over s^{d-2}} {\cal N}(s)\ 
\sim\ {1\over d-1} {N^2 V_{d-2}\over \epsilon^{d-2}}\ +\
N^2 \int {ds\over s} {V_{d-2}\over s^{d-2}} {s^{d-1}\over (1/T)^{d-1}} 
\Big|_l  \nonumber\\
&\sim& {1\over d-2} {N^2 V_{d-2}\over \epsilon^{d-2}}\ +\ 
N^2 T^{d-1} V_{d-2} l\ .
\eea
The energy enhancement factor ${s^{d-1}\over (1/T)^{d-1}}$ changes the 
IR behaviour as expected. The finite part of entanglement entropy is 
dominant for sufficiently large $l$ and is essentially the thermal
entropy in this regime. The linear growth with $l$ of the entropy
which is extensive is equivalent to the number of partons 
${\cal N}(s)$ being extensive.

For the nonconformal theory in $d=p+1$ dim at finite temperature $T$, 
with $\rho={E\over V}$ being the energy density, the thermal entropy 
$S(\rho, V)$ and temperature ${1\over T}={\del S\over \del E}$ are
\cite{Itzhaki:1998dd}
\be 
S \sim\  V g_{YM}^{(p-3)/(5-p)} \sqrt{N} \rho^{(9-p)/(2(7-p))}\ ,
\qquad
\rho\ \sim\ g_{YM}^{2(p-3)/(5-p)} N^{(7-p)/(5-p)}\ T^{2(7-p)/(5-p)}\ .
\ee
These can be recast as \cite{Barbon:2008ut}
\be
S \sim\ N_{eff}(1/T) V T^p ,\quad \rho \sim\ N_{eff}(1/T) T^{p+1} ,
\qquad 
N_{eff}(1/T) = N^2  (g_{YM}^2N T^{p-3})^{{p-3\over 5-p}}\ .
\ee
Along the lines earlier, we could obtain the total entanglement by 
integrating the number of entangling partons over length scales longer 
than that set by the temperature: this gives ($d=p+1$)
\be
S^{finite}\ \sim\ \int {ds\over s} {V_{d-2}\over s^{d-2}} 
N_{eff}(1/T)\ \Big({s\over (1/T)}\Big)^{d-1} \ 
\sim\ V_{d-2}l T^{d-1} N_{eff}(1/T)\ .
\ee
It is important to note that the thermal entropy is essentially 
classical, with contributions from partons of size predominantly 
${1\over T}$ so that we do not integrate $N_{eff}(s)$ over all scales 
$s$: \ie\ $N_{eff}=N_{eff}(1/T)$ above. In fact 
integrating the number of nonabelian degrees of freedom $N_{eff}(s)$ 
over scales $\epsilon < s < l$ in the above thermal context does not 
yield sensible results\ (\eg\ giving logarithmic growth for the 
thermal entropy for $p=1, 4$), in contrast with the ground state.

Now we want to interpret entanglement entropy for the pure CFT excited 
states dual to $AdS$ plane waves within this scaling picture. The 
energy density $T_{++}=Q$ sets a characteristic length scale $Q^{-1/d}$:
then the typical size of the partons is $Q^{-1/d}$~. Thus for cells 
of size $s$ much smaller than $Q^{-1/d}$, the parton distribution is 
similar to that in the ground state while for cells of size $s$ much 
larger than $Q^{-1/d}$, there is an enhancement in the number of 
entangling partons per cell. The anisotropy induced by the flux 
which is along one of the spatial directions implies that the 
entangling partons have energy-momentum in that direction but can be
regarded as essentially static in the other directions, as in the 
ground state. Consider first the case when the strip is along the 
flux direction: then as the strip width increases, the number of 
partons straddling the boundary increases since the partons move 
along the boundary. On the other hand, when the strip is orthogonal 
to the flux, the parton motion is orthogonal to the boundary: thus 
when the strip width is much larger than the characteristic size 
$Q^{-1/d}$ of the partons, the number of partons straddling the 
boundary is essentially constant since most of the partons enter 
the strip at one boundary and then shortly do not straddle the 
boundary but are completely encompassed within the strip. This 
reflects in the entanglement saturating for large width, with
the strip orthogonal to the energy flux.

Now we consider the case of the strip along the flux in more detail.
We again define the number of entangling bits or partons ${\cal N}(s)$
at scale $s$, with ${\cal N}(s)|_{s\ll Q^{-1/d}} \sim N^2$ for length
scales much smaller than the characteristic length $Q^{-1/d}$: above 
this scale, we expect some nontrivial scaling of ${\cal N}(s)$ 
which will be a function of $Qs^d$ on dimensional grounds. The precise 
functional form of ${\cal N}(s)$ for these $AdS$ plane wave states 
is not straightforward to explain however: the known results for 
holographic entanglement entropy (\ref{EEfinAdspw}) suggest\ 
${\cal N}(s) \sim\ N^2 \sqrt{Qs^d}$. Although the $AdS$ plane wave 
CFT states are simply the thermal CFT state in a low temperature large 
boost limit, this scaling of ${\cal N}(s)$ is not a simple boosted 
version of those for the thermal state (discussed below), but somewhat 
nontrivial. It would be interesting to explain this scaling of the 
$AdS$ plane wave CFT states, perhaps keeping in mind the infinite 
momentum frame and Matrix theory.  In this regard, we note that these 
$AdS$ plane wave states preserve boost invariance, \ie\ 
$x^\pm\ra \lambda^{\pm 1} x^\pm ,\  Q\ra \lambda^{-2}Q$ is a symmetry of 
the bulk backgrounds. For the strip along the flux, the longitudinal 
size scales as\ $V_{d-2}\ra \lambda V_{d-2}$ and the number of 
entangling partons is some function $f(Qs^d)$. 
Boost invariance then fixes $V_{d-2} f(Qs^d) = V_{d-2} \sqrt{Qs^d}$.\ 
Alternatively, imagine the collision of two identical plane wave 
states, moving in opposite directions. Assuming the resulting state 
has a number of partons\ ${\cal N}_L(s) {\cal N}_R(s)\propto Qs^d$ 
proportional to the energy-momentum density, we can estimate that 
either individual wave has ${\cal N}_L(s) \sim {\cal N}_R(s) \sim 
\sqrt{Qs^d}$. However this is a bit tricky since this makes 
${\cal N}_L(s), {\cal N}_R(s)$ reminiscent of partition functions: 
a number of partons might instead be expected to be additive, as 
${\cal N}_L(s)+ {\cal N}_R(s)$.

Taking the number of entangling partons ${\cal N}(s)$ at the boundary 
at scale $s\gg Q^{-1/d}$ as\ $N^2 {V_{d-2}\over s^{d-2}} \sqrt{Qs^d} = 
N^2 {V_{d-2}\over s^{d-2}} \left({s\over Q^{-1/d}}\right)^{d/2}$,
while for $s\ll Q^{-1/d}$ keeping $N^2 {V_{d-2}\over s^{d-2}}$ as in 
the ground state, gives rise to an entanglement scaling as
\bea
S &\sim& \int_\epsilon^l {ds\over s} {V_{d-2}\over s^{d-2}} {\cal N}(s)\ 
\sim\ {1\over d-2} {N^2 V_{d-2}\over \epsilon^{d-2}}\ +\
N^2 V_{d-2} \int {ds\over s} {\sqrt{Qs^d~} \over s^{d-2}}\Big|_l \nonumber\\
&\sim&\ \ {1\over d-2} {N^2 V_{d-2}\over \epsilon^{d-2}}\ +\ 
{N^2\over 4-d} \sqrt{Q}\ V_{d-2} l^{2-{d\over 2}}\ \qquad\quad [d\neq 4]\ ,
\nonumber\\
&\sim&\ \ {1\over d-2} {N^2 V_{d-2}\over \epsilon^{d-2}}\ +\ 
N^2 \sqrt{Q}\ V_{2} \log(lQ^{1/4}) \qquad\quad [d=4]\ .
\eea
For $d=4$, the logarithmic growth in the finite part arises by 
integrating from scales longer than $Q^{-1/4}$ upto the IR scale $l$.
Thus we see that the phenomenological scaling $\sqrt{Qs^d}$ is 
consistent with the holographic results. It would be interesting 
to understand this scaling better.\ 
Likewise for the nonconformal plane wave excited states (as in the 
conformal case) which we think of as chiral subsectors, the number of 
entangling partons at length scales $s$ longer than that set by the 
energy density $Q$ is proportional to $\sqrt{Qs^d}$ and the total 
finite part of entanglement for a strip subsystem of width $l$ becomes
\be
S_{finite} \sim\ \int {ds\over s}  {V_{d-2}\over s^{d-2}} 
\sqrt{N_{eff}(s)}\ \sqrt{Qs^d~}\Big|_l \ \sim\  {5-p\over 3-p} 
{V_{d-2}\over l^{d-2}}\ \sqrt{N_{eff}(l)}\  \sqrt{Ql^d}\ ,
\ee
recovering the holographic results \cite{Narayan:2013qga}.\\
It would be interesting to put the phenomenological discussions in 
this section on firmer footing with a view to gaining deeper insight 
into entanglement in field theory excited states.

\section{Holographic mutual information: generalities}

\begin{figure}[h]
\bc
\includegraphics[width=30pc]{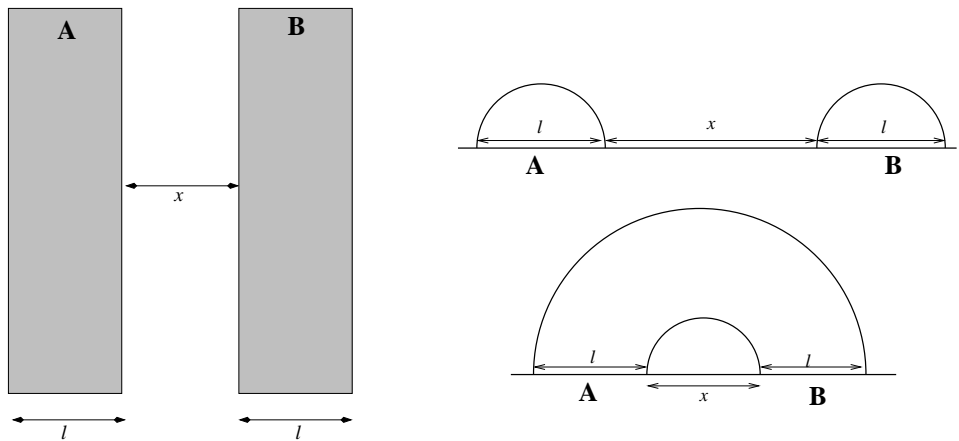} \ec  
\caption{{\label{figMItransition}\footnotesize { 
Two parallel disjoint strip subsystems of width $l$ and separation $x$ 
(and longitudinal size $V_{d-2}$) (left), with the disconnected extremal 
surface (top right) and the connected extremal surface (bottom right).
}}}
\end{figure}
Mutual information is defined for two disjoint subsystems $A$ and $B$ as
\be
I[A,B] = S[A] + S[B] - S[A\cup B]\ .
\ee
It is a measure of the correlation (both classical and quantum)
between the degrees of freedom of two disjoint subsystems $A$ and
$B$. Mutual information is finite,\ positive semi-definite, and 
proportional to entanglement entropy when $B \equiv A^c$ (in that 
case, $S(A \cup A^c)=0$). 
This linear combination of entanglement entropies ensures that the
short distance area law divergence cancels between the various
individual terms rendering the mutual information finite. There is a
new cutoff-independent divergence however that arises when the two 
subsystems approach each other and collide, as we will see below.

The holographic prescription of Ryu-Takayanagi implies in a simple
geometric way that mutual information vanishes when the two subsystems
are widely separated: thus as discussed in \cite{Headrick:2010zt},
mutual information undergoes a disentangling phase transition as the
separation between the two striplike subsystems $A$ and $B$ increases.
Recall that we choose that extremal surface which has minimal area,
given the boundary conditions defined by the subsystem in question.
In the case of the subsystem $A\cup B$ defined by two disjoint strips,
there are two candidate extremal surfaces as in
Figure~\ref{figMItransition}. When the two subsystems are widely
separated, the relevant extremal surface with lower area is simply the
union of the two disconnected surfaces so that $S[A\cup B] = S[A] +
S[B]$. However for nearby subsystems, the connected surface has lower
area.  For simplicity, we consider two disjoint parallel strip
subsystems with longitudinal size $V_{d-2}$, and of the same width $l$
each, with separation $x$. For fixed width $l$, we can vary the
separation $x$. Then as we vary ${x\over l}$ which is a dimensionless
parameter, the behaviour of the extremal surface and its area $S[A\cup
B]$ change: the extremal surface is
\bea\label{MIgenstruc}
&(i)& \mbox{the disconnected surface:\ area}\ \ S[A\cup B] = 
S(A) + S(B) = 2S(l)\ ,\ \ 
\mbox{for large}\ \ {x\over l}\ ,\nonumber\\ 
&(ii)& \mbox{the connected surface:\ area}\ \ S[A\cup B]= S(2l+x)+S(x)\ ,
\ \ \mbox{for small}\ \ {x\over l}\ .
\eea
The Ryu-Takayanagi prescription of choosing the extremal surface of
minimal area then leads to a change in the entangling surface for the
combined subsystem $A\cup B$. Correspondingly the mutual
information changes as 
\be
\begin{aligned}
 I[A,B]\ & >\ 0, \qquad\qquad\qquad {x\over l}\ <\ {x_c\over l}\ ,\\
       \ & =\ 0, \qquad\qquad\qquad {x\over l}\ >\ {x_c\over l}\ .
\end{aligned}
\ee
The critical value ${x_c\over l}$ is a dimensionless number, and depends 
on the field theory in question as well on the CFT state, as we discuss
below. This critical value ${x_c\over l}$ is thus the location of a 
sharp disentangling transition in the classical gravity approximation, 
since the mutual information vanishes for larger separations implying 
the subsystems are uncorrelated, especially in light of an interesting 
relation between the mutual information and correlation functions.
It is known \cite{wolfEEmi} that $I[A,B]$ sets an upper bound for 
2-point correlation functions of operators, with one insertion at a 
point in region $A$ and the other in $B$,
 \be
 \label{MIandcorr}
 I[A,B] \geq \frac{(\lA \cO_A \cO_B\rA -\lA \cO_A\rA \lA \cO_B\rA)^2}
{2|\cO_A|^2|\cO_B|^2} \ .
\ee
This inequality implies that beyond the disentangling transition point 
all 2-point correlation functions also vanish (with one point in $A$ 
and the other in $B$), since the mutual information vanishes. It is 
important to note that entanglement entropy and mutual information via 
the Ryu-Takayanagi prescription are $O(N^2)$ observables in the classical 
gravity approximation. However the 2-pt correlators are normalized as 
$O(1)$. One might imagine the mutual information decays as 
$I[A,B]\sim \sum {c_\Delta\over r_{A,B}^{4\Delta}}$ and indeed the quantum 
$O(1)$ contributions effectively give a long distance expansion for 
mutual information \cite{Faulkner:2013ana} (see also \cite{Headrick:2010zt} 
\cite{VanRaamsdonk:2009ar} \cite{Huijse:2011ef} \cite{Casini:2008wt}). 
However the coefficients $c_\Delta$ at the classical level $O(N^2)$ 
vanish: this shows up in the large $N$ approximation as the sharp 
disentangling transition in mutual information.

We now discuss this for large $N$ conformal field theories in the 
ground and excited states.
For the ground state, the mutual information for two strip shaped 
subsystems of width $l$ parallel to each other and with separation 
$x$ is
\be\label{miGndSt}
I[A,B]\ =\ -c V_{d-2} \left( {2\over l^{d-2}} - {1\over (2l+x)^{d-2}} 
- {1\over x^{d-2}} \right) 
=\ -c {V_{d-2}\over l^{d-2}} \left( 2 - {1\over (2+{x\over l})^{d-2}} 
- {1\over ({x\over l})^{d-2}} \right) .
\ee
This arises from the cutoff-independent parts of entanglement, the 
divergent terms cancelling. We see that for small separation $x$, the 
mutual information $I[A,B]$ grows as $I[A,B]\sim {V_{d-2}\over x^{d-2}}$ 
and exhibits a divergence as $x\ra 0$, \ie\ when the subsystems collide 
with each other. As ${x\over l}$ increases, $I[A,B]$ decreases and then 
vanishes at a critical value of ${x\over l}$. Beyond this critical 
separation, the expression (\ref{miGndSt}) for $I[A,B]$ as it stands 
is negative and is meaningless: this simply reflects the fact that 
the correct extremal surface for $A\cup B$ is in fact the disconnected 
surface, \ie\ the subsystems disentangle, and mutual information 
actually vanishes beyond the disentangling point.
This disentangling transition can be identified as the zero of $I[A,B]$ 
above, giving\  ${x_c\over l} \simeq 0.732\ [d=4]$, and $0.62\ [d=3]$, 
and so on.

Such a disentangling transition also happens at finite temperature, but 
the phase diagram is more complicated and has nontrivial dependence on 
the length scale ${1\over T}$ set by the temperature $T$. For 
$l, x \ll {1\over T}$, \ie\ subsystem widths and separation small 
relative to the temperature scale, we only expect small corrections 
to the ground state behaviour above. Thus the disentangling transition 
point occurs at values which are ``near'' those for the ground state.
However for large width $l$, the entanglement is well approximated by 
the extensive (linear) thermal entropy: thus 
\be
I[A,B] \sim\ T^{d-1} V_{d-2}\ \left(2l - (2l+x) - S^{fin}(x)\right) 
=\ T^{d-1} V_{d-2}\ \left(- S^{fin}(x)\ -\ x\right).
\ee
Thus we see that as the separation $x$ increases, $-S^{fin}(x)>0$ 
decreases and $I[A,B]$ decreases and eventually vanishes at a 
critical $x_c$, which turns out to be smaller in value than for the 
ground state. When the thermal entropy dominates the entanglement or 
equivalently the subsystem widths and separation are both large 
relative to the temperature scale, we see that 
\be\label{MIdisentFiniteT}
I[A,B] \sim\ T^{d-1} V_{d-2}\ \left(2l - (2l+x) - x\right) 
= -2 T^{d-1} V_{d-2} x\ ,
\ee
which is negative. This is a reflection of the fact that the two 
subsystems in fact are completely disentangled for any separation 
$x$ larger than ${1\over T}$. In some sense, the temperature 
``disorders'' the system and the subsystems disentangle faster at 
finite temperature than in the ground state.

In what follows, we analyse holographic mutual information for $AdS$ 
plane waves. We will see some similarities with the finite 
temperature case, but with nontrivial phase structure depending on 
the scale $Q^{-1/d}$. There are however some key differences 
as we will see below.

\section{Mutual information in $AdS$ plane waves}

$AdS$ plane waves exhibit anisotropy due to the energy flux in one
direction. We are considering parallel disjoint strip subsystems that
are either both along the flux or both orthogonal to the flux. We can
analyse mutual information in two extreme regimes, where the strip
widths $l$ are large or small compared to the length scale set by the
energy density flux $Q$. Eventually we will carry out some numerical 
analysis in intermediate regimes as well.

\subsection{Wide strips:\ $l Q^{1/d} \gg 1$}

Consider first the strip along the energy flux direction, with width 
direction along say $x_1$\ (we assume $d\geq 3$). Then the spacelike 
strip subsystem $A$ lying on a constant time slice has 
$0\leq x_1\leq l,\ (x^+,x^-) = (\al y,-\beta y) = (y,-y),\ 
-\infty<y,x_2,x_3,\cdots,x_{d-2}<\infty$.
The extremal surface $\gamma_A$ is specified by the function $x_1=x(r)$. 
$V_{d-2}$ denotes the volume in the $y$ and $(x_2, \cdots,x_{d-2})$ 
direction. $\epsilon$ is the UV cutoff. 
The subsystem width in terms of the turning point $r_*$ is 
\cite{Narayan:2012ks}
\be\label{caseawidth}
\Delta x_1 = l=2\int^{r_*}_0 dr \frac{Ar^{d-1}}{\sqrt{2+Qr^d-A^2r^{2(d-1)}}}\ ,
\ee
while the entanglement entropy in terms of the area functional is 
\be\label{caseaent}
S_A = {Area\over 4G_{d+1}} = 
{2V_{d-2}R^{d-1}\over 4G_{d+1}}\int^{r_*}_\epsilon \frac{dr}{r^{d-1}}
\frac{2+Qr^{d}}{\sqrt{2+Qr^d-A^2 r^{2(d-1)}}}\ .
\ee
There is a leading area law divergence from the contribution near the 
boundary $r=\epsilon$, with\ 
$EE \sim\ N^2 {V_{d-2}\over\epsilon^{d-2}}$~,\ where we have used 
$N^2\sim {R^{d-1}\over G_{d+1}}$.
For large energy density $Q$, and large width $l$, the turning point 
equation $2+Qr_*^d-A^2r_*^{2d-2}=0$ can be approximated as\ 
$Qr_*^d\simeq A^2r_*^{2(d-1)}\gg 1$, so that\ $l\sim r_*$ from 
(\ref{caseawidth}). The finite cutoff-independent piece of $S_A$ is 
then estimated as
\bea\label{eesl}
S_A^{finite}&\sim& \pm {R^{d-1}\over G_{d+1}} V_{d-2}\sqrt{Q}\ 
l^{2-\frac{d}{2}}\qquad\ \ 
[d\neq 4]\\
&\sim& N^2 V_{2}\sqrt{Q}\ \log (l Q^{1/4} )\qquad\ \ [d=4]\ . \label{loge}
\eea
The sign in front of (\ref{eesl}) is $+$ for $d<4$ and $-$ for $d>4$.

Towards estimating mutual information for $AdS$ plane waves, we must 
note that there are multiple regimes stemming from the various length 
scales $l,\ x,\ Q^{-1/d}$. When the strip widths and separations are 
large relative to the correlation length, \ie\ $lQ^{1/d}\gg 1$ and 
$xQ^{1/d}\gg 1$, we can use the above estimates for the finite parts 
of entanglement entropy to estimate mutual information. For the 
$AdS_5$ plane wave, when the strips are not too far apart, we can 
assume mutual information is nonzero, obtaining from the finite parts 
above,
\be\label{MIlargeQld}
I[A,B] = 2S^{fin}(l)-S^{fin}(2l+x)-S^{fin}(x) 
\sim\ V_2\sqrt{Q} \log \left({l^2\over x(2l+x)}\right) .
\ee
The argument of the logarithm vanishes when
\be
I[A,B]\ra 0\qquad \Rightarrow\qquad l^2=2lx+x^2,\qquad\ie\ \ \
{x_c\over l}=\sqrt{2}-1\simeq 0.414\ .
\ee
Thus the subsystems disentangle at a separation less than that for the
$AdS_5$ ground state, which has ${x_c\over l}=0.732$. It is also
noteworthy that for any large $Q$, the subsystems disentangle only
when they are sufficiently wide apart in comparison with the width,
\ie\ $x\geq 0.414 l$, independent of the characteristic energy scale
$Q^{-1/4}$: in particular the disentangling point $x_c$ here could be
substantially bigger than $Q^{-1/4}$. This transition location agrees
with the analysis for hyperscaling violating spacetimes in
\cite{Huijse:2011ef} and \cite{Fischler:2012uv}, in accordance with
the fact that the $AdS_5$ plane wave gives rise to the corresponding
hyperscaling violating spacetime. The strips, being parallel to the
flux, are unaffected by the reduction along the $x^+$-circle from that
perspective. In the present case, we are studying this entirely from
the higher dimensional $AdS$ plane wave point of view.
Note that this is quite distinct from the finite temperature case 
\cite{Fischler:2012uv} in the corresponding regime $lT\gg 1,\ xT\gg 1$, 
\ie\ sizes larger than the temperature scale ${1\over T}$~: in that 
case, the linear extensive growth of entanglement in this regime 
implied that the subsystems disentangled for any finite separation 
$x$ independent of the width $l$ (\ref{MIdisentFiniteT}).

Strictly speaking, we are thinking of the regulated $AdS$ plane wave
as a limit of the highly boosted low temperature $AdS$ black brane,
with a large separation of scales $Q\gg r_0^4$ between the energy
density $Q=\lambda^2r_0^4$ and the temperature $r_0$, with $\lambda$
being the boost parameter. Over this wide range of length scales, the
physics is dominated by the $AdS$ plane wave description, with
departures arising in the far infrared where the black brane horizon
physics enters as a regulator.  From this point of view, we are
thinking of the strip subsystem widths as satisfying $Q^{-1/4}\ll l\ll
{1\over r_0}$, with the above behaviour of mutual information holding
correpondingly: in the far IR when $l\gg {1\over r_0}$ the behaviour
of mutual information resembles that in the finite temperature case.

A similar analysis can be done for the $AdS_4$ plane wave, in the 
regime $lQ^{1/3} \gg 1$ and $xQ^{1/3}\gg 1$, taking again for simplicity 
both strips of 
equal width $l$ with separation $x$. Then the mutual information arises 
from the finite parts of entanglement estimated (\ref{eesl}) for large 
$Q$ giving
\be
I[A,B] \sim\ V_1 \sqrt{Q} \left( 2\sqrt{l} - \sqrt{2l+x} - \sqrt{x}\right) .
\ee
This decreases as the separation $x$ increases and finally vanishes 
when 
\be
I[A,B] \ra 0\qquad \Rightarrow\qquad {x_c\over l} = {1\over 4}\ ,
\ee
which is the location of the disentangling transition in this regime.
Again we see that the subsystems disentangle when they are sufficiently 
wide apart in comparison to their widths $l$, without specific dependence 
on the energy scale $Q^{-1/3}$ as for the $AdS_5$ plane wave discussed 
above.

Nonconformal D-brane plane waves and entanglement entropy were studied 
in \cite{Narayan:2013qga}, with the emerging picture and scalings 
consistent with $AdS$ plane waves in cases where comparison is possible. 
The analysis is more complicated in the nonconformal cases since there 
are multiple different length scales in the phase diagram. 
The structure of mutual information is still further complicated and we 
will not carry out a systematic study here. We can however make some
coarse estimates in the large flux regime. For instance the D2-M2 
ground state phase diagram \cite{Itzhaki:1998dd} extends to a 
corresponding one for the D2-brane plane waves. The finite part of 
EE for a strip along the flux in the D2-brane supergravity regime is\ 
$S^{fin}_{D2}\sim\ V_1\sqrt{Q} \sqrt{l} \sqrt{N_{eff}(l)}\sim\ 
V_1\sqrt{Q} \sqrt{l} \sqrt{{N^2\over (g_{YM}^2Nl)^{1/3}}}\propto l^{1/3}$. 
Noting the D2-sugra regime of validity, it can be seen that this 
finite part is greater than $V_1\sqrt{Q} \sqrt{l} \sqrt{N^{3/2}}$ 
for the M2-brane ($AdS_4$) plane wave arising in the far IR 
\cite{Narayan:2013qga}. In the D2-regime, we can approximate the 
mutual information as\
$MI_{D2}\sim\ V_1\sqrt{Q} (2l^{1/3}-(2l+x)^{1/3}-x^{1/3})$ which shows 
a disentangling transition at ${x_c\over l}\sim 0.31$. Recalling that 
for the M2-brane regime, we have ${x_c\over l}\sim 0.25$, we see 
that $x_c$ decreases along the RG flow from the D2-brane sugra to 
the M2-brane regime. Similarly for the ground states also, it can 
be checked that in the D2-regime, we have ${x_c\over l}\sim 0.66$ 
while in the M2-regime, we have ${x_c\over l}\sim 0.62$. It is 
unclear if these are indications of some deeper structure for the 
``flow'' of mutual information.

Now we make a few comments on mutual information in the case where 
the strips are orthogonal to the energy flux. In the large flux regime, 
we know \cite{Narayan:2012ks} that entanglement entropy shows a 
phase transition for $l\gg Q^{-1/d}$ with no connected extremal 
surface but only disconnected ones. In this regime, we expect that 
mutual information simply vanishes since the connected surface of 
mutual information (\ref{MIgenstruc}) is already disconnected: thus 
the entanglement is saturated for each of $S[l], S[2l+x], S[x]\sim 
S_{sat}$ so that $MI\sim 2S(l)-S[2l+x]-S[x] =0$. In sec.~5.3, we 
will study entanglement and mutual information in the perturbative 
regime $Ql^d\ll 1$: however in this regime, we do not expect any 
signature of the phase transition which is only visible for wide 
strips. It is then reasonable to expect some interesting interplay 
between the phase transition and the location of the disentangling 
transition for mutual information.

\subsection{Narrow strips: $l Q^{1/d} \ll 1$, \ strips along flux}

We would now like to understand the case of narrow strips, \ie\ with 
the dimensionless quantity $lQ^{1/d}\ll 1$. In this limit, we expect 
that the entanglement entropy is only a small departure from the pure 
$AdS$ case, since the energy density flux $Q$ will only make a small 
correction to the ground state entanglement. We will first analyse the 
strip along the flux and obtain the entanglement correction to the 
ground state. This has parallels with ``entanglement thermodynamics'' 
\cite{Bhattacharya:2012mi} \cite{Allahbakhshi:2013rda} 
\cite{Wong:2013gua} for these $AdS$ plane waves, treating the $g_{++}$ 
mode as a small deformation to $AdS$.

In the limit $Q^{1/d}l\ll 1$, we first calculate the change 
in the turning point $r_*$ upto $O(Q)$, and then expand the width 
integral and area integral around $AdS_{d+1}$, using (\ref{caseawidth}), 
(\ref{caseaent}).
First we note that the pure $AdS$ case, with $s$ the turning point of 
the minimal surface, has the width integral
\be\label{pureadswidth}
l = 2\int_0^s\frac{A}{\sqrt{\frac{2}{r^{2(d-1)}}-A^2}} = 
2\int_0^s dr \frac{(r/s)^{d-1}}{\sqrt{1-\left(\frac{r}{s}\right)^{2(d-1)}}}
= 2 \left(\frac{\sqrt{\pi}\Gamma\left(\frac{d}{2d-2}\right)}{\Gamma\left(\frac{1}{2d-2}\right)}\right) s\ \equiv\ 2 \eta s\ ,
\ee
using $A^2 = \frac{2}{s^{d-1}}$ and 
$\eta = \int_0^1 \frac{x^{d-1}}{\sqrt{1-x^{2(d-1)}}}dx$.
We want to calculate the change in the ground state entanglement 
entropy under the $AdS$ plane wave perturbation to $O(Q)$, with the 
strip along the flux. With the entangling surface fixed at width $l$, 
the turning point $s$ now changes to $r_*=s+\delta r_*$. 
We recast (\ref{caseawidth}) and the turning point equation as
\be\label{AQr*}
 \frac{l}{2} =\int_0^{r_*} dr \frac{A}{\sqrt{\frac{g(r)}{r^{2(d-1)}}-A^2}} 
\quad \mbox{with}  \quad g(r)=2+Qr^d\ ,\quad \mbox{and}  \quad 
A^2 = \frac{g(r_*)}{r_*^{2(d-1)}} \equiv \frac{g_*}{r_*^{2(d-1)}}\ .
\ee
Then we obtain
\be
\frac{l}{2} = \int_0^{r_*}dr\frac{\frac{\sqrt{g_*}}{r_*^{d-1}}}
{\frac{1}{r^{d-1}}\sqrt{g(r)-{g_*}\left(\frac{r}{r_*}\right)^{2(d-1)}}}\ 
=\ \int_0^{r_*}dr\frac{(r/r_*)^{d-1}}{f(r,r_*)}
\frac{\left(1+\frac{Qr_*^d}{4}\right)}
{\sqrt{1+\frac{Qr^d-Qr_*^d\left(\frac{r}{r_*}\right)^{2(d-1)}}{2f^2(r,r_*)}}}\ ,
\ee
with the function
\be\label{frr*}
f(r,r_*)=\sqrt{1-\left(\frac{r}{r_*}\right)^{2(d-1)}}\ ,\qquad\quad
0<f(r,r_*)<1 \quad\ \mbox{for all}\ r<r_*\ .
\ee
The above expression has been obtained by taking $Qr_*^d\ll 1$ and 
expanding out the integrand. 
The above width integral can be further simplified to $O(Q)$ as 
\bea
\frac{l}{2} &=& \int_0^{r_*}dr\frac{(r/r_*)^{d-1}}{f(r,r_*)}
{\left(1+\frac{Qr_*^d}{4}\right)} 
\left({1-\frac{Qr^d-Qr_*^d\left(\frac{r}{r_*}\right)^{2(d-1)}}{4f^2(r,r_*)}}\right) \nonumber\\
&=& \int_0^{r_*}dr\frac{(r/r_*)^{d-1}}{f(r,r_*)}
\left(1+\frac{Q}{4f^2(r,r_*)}(r_*^d-r^d)\right) =\ 
s\eta = (r_*-\delta r_*) \eta\ ,
\eea
the last expression arising since the width $l$ is as in $AdS$.
Using (\ref{pureadswidth}), we see that the leading $AdS$ piece cancels 
giving
\be\label{caseacorr}
\delta r_* = -\frac{Q}{4\eta}\int_0^{r_*}dr 
\frac{(r/r_*)^{d-1}}{f^3(r,r_*)}\Big(r_*^d-r^d\Big) \sim\ 
-\frac{Qs^{d+1}}{4\eta} \int_0^1 dx 
\frac{x^{d-1}(1-x^d)}{(1-x^{2(d-1)})^{3/2}}\ .
\ee
As $r_*$ happens to be the turning point of the minimal surface, 
$r< r_*$ which implies that $\delta r_*<0$ always. Also since $\delta r_*$ 
is $O(Q)$, we have approximated $r_*\sim s$ to obtain the second expression. 
Thus
\be\label{caseacorr2}
\delta r_*  \sim\ -\frac{Qs^{d+1}}{4\eta}\frac{\sqrt{\pi}}{(d-1)^2}
\left(\frac{\Gamma(\frac{1}{d-1})}{\Gamma(\frac{1}{2}+\frac{1}{d-1})}
-(d-1)\frac{\Gamma(\frac{d}{2d-2})}{\Gamma(\frac{1}{2d-2})}\right) 
\equiv\ -\frac{Qr_*^{d+1}}{4\eta} {\cal N}_{r_*}\ .
\ee
We now calculate the change in the area integral and correspondingly 
the entanglement entropy upto $O(Q)$. For pure $AdS$, \ie\ the CFT 
ground state, we have
\be\label{pureadsee}
4G_{d+1} S_0 = 2V_{d-2}R^{d-1}\int_0^s \frac{dr}{r^{d-1}}\frac{1}{f(r,s)}\ ,
\ee
with $f(r,s)= \sqrt{1-\left(\frac{r}{s}\right)^{2(d-1)}}$ as 
in (\ref{frr*}).
We focus on the finite part of the above integral and use 
$l=2s\eta$, obtaining
\be\label{pureAdSEE2}
4G_{d+1} S_0\ =\ \# R^{d-1} {V_{d-2}\over\epsilon^{d-2}}\ -\ 
\frac{2^{d-1}\pi^{\frac{d-1}{2}}}{(d-2)}
\left(\frac{\Gamma(\frac{d}{2d-2})}{\Gamma(\frac{1}{2d-2})}\right)^{d-1} 
{V_{d-2}\over l^{d-2}} R^{d-1}\ .
\ee
In our case of the $AdS_{d+1}$ plane wave,
\be
4G_{d+1} S = 2V_{d-2}R^{d-1}\int_0^{r_*}\frac{dr}{r^{d-1}}
\frac{2+Qr^d}{\sqrt{2+Qr^d-A^2r^{2(d-1)}}}
\ee
Treating this as an infinitesimal $g_{++}$-deformation and expanding
around pure $AdS$, we would like to obtain the $O(Q)$ change in EE, 
or equivalently the infinitesimal change for the plane wave excited 
state relative to the ground state. From the turning point equation, 
we have $A^2 = \frac{2+Qr_*^d}{r_*^{2(d-1)}}$ as before, giving
\bea
4G_{d+1} S &=& 2V_{d-2}R^{d-1}\int_0^{r_*}\frac{dr}{r^{d-1}}
 \frac{2+Qr^d}{\sqrt{2\Big(1-\Big(\frac{r}{r_*}\Big)^{2(d-1)}\Big)
 +Qr^d-Qr_*^d\Big(\frac{r}{r_*}\Big)^{2(d-1)}}} \nonumber\\
&=& 2\sqrt{2}V_{d-2}R^{d-1}\int_0^{r_*}\frac{dr}{r^{d-1}}\frac{1}{f(r,r_*)}
 \left(1+\frac{Qr^d}{2}\right)\left(1-\frac{Qr^d-Qr_*^d\left(\frac{r}{r_*}
\right)^{2(d-1)}}{4f(r,r_*)^2}\right)  \nonumber\\
&=& 4G_{d+1} S_0\ +\ 2\sqrt{2}R^{d-1} {\cal N}_{EE}\ V_{d-2} Qr_*^2\ ,
\eea
where 
\be
\begin{aligned}
{\cal N}_{EE} &=\int_0^1 dx\left[\frac{x}{2\sqrt{1-x^{2(d-1)}}} +
\frac{1}{4x^{d-1}\sqrt{1-x^{2(d-1)}}}
\left(\frac{(1-x^d)} {(1-x^{2(d-1)})}-1\right)\right]\\
\ &=\ \frac{\sqrt{\pi}}{8(d-1)^2}\left(\frac{(d+1)\Gamma(\frac{1}{d-1})}
{\Gamma(\frac{1}{2}+
\frac{1}{d-1})}-\frac{2(d-1)\Gamma(\frac{d}{2d-2})}
{\Gamma(\frac{1}{2d-2})}\right)\ .
\end{aligned}
\ee
It can be checked that the constant ${\cal N}_{EE}$ is positive, so that the 
correction to the entanglement entropy is positive. To $O(Q)$, we can 
replace $r_*$ by $s$, the pure $AdS$ turning point. Then using 
$l=2s\eta$, we see that
\be
\Delta S  \sim\ + {R^{d-1}\over G_{d+1}} {{\cal N}_{EE}\over 4\eta^2\sqrt{2}}\ 
V_{d-2} Q l^2\ = +{R^{d-1}\over G_{d+1}} {{\cal N}_{EE}\over 4\eta^2\sqrt{2}}\ 
{V_{d-2}\over l^{d-2}} (Q l^d)\ ,
\ee
with $Ql^d\ll 1$.\ There are parallels of this analysis with 
``entanglement thermodynamics'' \cite{Bhattacharya:2012mi,
Allahbakhshi:2013rda,Wong:2013gua} (see also 
\cite{Blanco:2013joa,Lashkari:2013koa,Faulkner:2013ica}). 
In the present case, we have the energy change in the strip\ 
$\Delta E \sim \int \delta T_{tt} d^{d-1}x \sim\ 
Q V_{d-2} l$, giving $T_E\Delta S_E \sim \Delta E$ with the 
``entanglement temperature'' $T_E\sim {1\over l}$~. There is also 
an entanglement pressure. Although it is not crucial for our purposes 
here, it would be interesting to develop this further.

The above entanglement entropy change implies that the change in 
mutual information is negative (with $I_0[A,B]$ the mutual 
information in pure $AdS$):
\bea\label{deltaMI1}
I[A,B] &=& I_0[A,B]\ +\ \Delta I[A,B]\ 
=\ I_0[A,B]\ +\ {R^{d-1}\over G_{d+1}} {{\cal N}_{EE}\over\sqrt{2}}\ 
V_{d-2} Q \left( 2l^2 - (2l+x)^2 - x^2 \right) \nonumber\\
&=& I_0[A,B]\ -\ 2 {R^{d-1}\over G_{d+1}} {{\cal N}_{EE}\over 4\eta^2\sqrt{2}}\ 
V_{d-2} Q l^2\ \left(1+{x\over l}\right)^2\ .
\eea
Thus we see that mutual information strictly decreases, for a small 
$T_{++}$ energy density flux perturbation along the strip subsystem.
In this perturbative regime with the correction scaling as $O(Q)$ and 
as the area of the interface $V_{d-2}$, the entanglement and mutual 
information corrections involve the dimensionless quantity 
$V_{d-2} Q l^2$.

It is worth noting that unlike in the wide strip regime 
(\ref{MIlargeQld}), the disentangling transition in this perturbative 
regime certainly depends on the energy density $Q$ and the strip width 
through $Ql^d$. In particular, using (\ref{pureAdSEE2}), (\ref{miGndSt}), 
we see that the mutual information (\ref{deltaMI1}) vanishes at 
\be
{\cal N}^0_{EE} \Big({1\over ({x\over l})^{d-2}} + 
{1\over (2+{x\over l})^{d-2}}-2\Big) 
- {{\cal N}_{EE}\over 2\sqrt{2}\eta^2} Ql^d \Big(1+{x\over l}\Big)^2 = 0\ ,
\ee
where ${\cal N}^0_{EE}$ is the constant coefficient of the finite part 
in (\ref{pureAdSEE2}). A numerical study later (sec. 5.4) describes 
the location of the vanishing of mutual information and the 
disentangling transition for intermediate regimes as well, where 
$Ql^d\sim O(1)$.

\subsection{Narrow strips: $l Q^{1/d} \ll 1$,\ strips orthogonal to flux}

We describe the change in entanglement entropy and mutual information 
for the strips orthogonal to the flux in the perturbative regime 
$l Q^{1/d} \ll 1$ here. The analysis is similar to the previous 
case, but involves more calculation.

We first consider a single strip and study entanglement. In this case, 
the width direction of the strip $A$ is parallel to $x_{d-1}$, with 
$x^{\pm} = \frac{t \pm x_{d-1}}{\sqrt{2}}$. The bulk extremal surface 
$\gamma_A$ is specified by $x^+=x^+(r),\ x^-=x^-(r)$, and the spacelike 
strip subsystem has width
\be\label{casebspacelike}
\Delta x^{+} = -\Delta x^{-} = \frac{l}{\sqrt{2}} > 0\ ,
\ee
(spacelike implying $\Delta t=0$) and longitudinal size 
$V_{d-2}\sim L^{d-2}$ with $L\gg l$ in the $x_i$ directions.
$\epsilon$ is the UV cut-off. Then the width integrals and the 
entanglement entropy area functional reduce to \cite{Narayan:2012ks}
\bea\label{casebwidth}
&& \frac{\Delta x^+}{2}=\int^{r_*}_0
\frac{dr}{\sqrt{\frac{A^2B^2}{r^{2(d-1)}}+Qr^{d}-2B}}\ ,\qquad 
\frac{\Delta x^-}{2}=\int^{r_*}_0
\frac{(Qr^d-B)\ dr}{\sqrt{\frac{A^2B^2}{r^{2(d-1)}}+Qr^{d}-2B}}\ ,\\
\label{SEadspw} && S_A\ =\ {2R^{d-1}V_{d-2}\over 4G_{d+1}}
\int^{r_*}_{\epsilon}\frac{dr}{r^{d-1}}
\frac{AB}{\sqrt{A^2B^2-2Br^{2(d-1)}+Qr^{3d-2}}}\ . 
\eea
Unlike the previous case, here we have two parameters $A, B$ and 
two integrals specifying the subsystem width $l$ as a function of the
turning point $r_*$ of the extremal surface, given by (\ref{casebwidth}). 
For pure $AdS$, with $Q=0$, (\ref{casebwidth}) alongwith 
(\ref{casebspacelike}) fixes $B=1$, with $x^\pm$ treated ``symmetrically'' 
as expected in the absence of the energy flux. We will treat the 
$AdS$ plane wave case in $O(Q)$ perturbation theory and expand 
both integrals around $AdS$. The turning point equation here is
\be
\frac{A^2B^2}{r_*^{2(d-1)}}+Qr_*^d -2B=0\qquad\Rightarrow\qquad
\frac{A^2B^2}{r^{2(d-1)}} = \left(\frac{r_*}{r}\right)^{2(d-1)}(2B-Qr_*^d)\ .
\ee
This recasts the denominator of the width integrals in terms of 
$f(r,r_*)=\sqrt{1-\left(\frac{r}{r_*}\right)^{2(d-1)}}$ and $B$ alone,
\be\label{denomDx+-}
\left[\frac{A^2B^2}{r^{2(d-1)}}+Qr^d-2B\right]^{1/2}
= \left(\frac{r_*}{r}\right)^{d-1}f(r,r_*) \sqrt{2B} 
\left[1-\frac{Qr_*^d\Big(1-\big(\frac{r}{r_*}\big)^{3d-2}\Big)}{2Bf^2}
\right]^{1/2} .
\ee
However unlike (\ref{AQr*}) earlier, we are still left with the 
parameter $B$ here, so the turning point equation does not suffice. 
The other relation for recasting both $A$ and $B$ in terms of $Q, r_*$ 
comes from the fact that we have a space-like subsystem, \ie\ 
(\ref{casebspacelike}). Specifically with the pure $AdS$ case 
corresponding to $B=1$, in this perturbative regime with $Ql^d\ll 1$, 
we can safely assume that $B = 1 + \delB$ with $\delB \sim O(Q)$. 
Since the two width integrals for $\Delta x^+$ and $\Delta x^-$ must 
obey the equality $\Delta x^+=-\Delta x^-={l\over\sqrt{2}}$, we 
must have that the change in the turning point $\delta r_*$ obtained 
from both is the same, which fixes $\delB\propto Qr_*^d$ as we will see.

To elaborate, from (\ref{casebspacelike}), (\ref{casebwidth}), 
(\ref{denomDx+-}), we have
\be\label{lDx+B}
\frac{\Delta x^+}{\sqrt{2}}\ =\ \frac{l}{2}\ =\ 
\int_0^{r_*}dr\frac{(r/r_*)^{d-1}}{f(r,r_*) \sqrt{B}} \frac{1}{\left[1 - 
\frac{Qr_*^d\left(1-(r/r_*)^{3d-2}\right)}{2Bf^2}\right]^{1/2}}\ .
\ee
Now, with $B=1+\delB=1+O(Q)$, we can expand this to $O(Q)$ obtaining
\be
\frac{l}{2}=\int_0^{r_*}dr\frac{(r/r_*)^{d-1}}{f(r,r_*)}
-\frac{\delB}{2}\int_0^{r_*}dr\frac{(r/r_*)^{d-1}}{f(r,r_*)}
+ Qr_*^d\int_0^{r_*}dr\frac{(r/r_*)^{d-1}(1-(r/r_*)^{3d-2})}{4f^3(r,r_*)}\ .
\ee
As in the previous subsection, we keep our entangling surface fixed 
so $l=2s\eta$, with $s$ the pure $AdS$ turning point. The new turning 
point is $r_* = s+\delta r_*$, so $l/2 = r_*\eta -\delta r_*
\eta$. Thus 
\be\label{deltar*x+}
-\delta r_* \eta = -\frac{\delB r_*}{2}\int_0^1 dx 
\frac{x^{d-1}}{\sqrt{1-x^{2(d-1)}}}
+Qr_*^{d+1}\int_0^1 dx \frac{x^{d-1}(1-x^{3d-2})}{4(1-x^{2(d-1)})^{3/2}}\ .
\ee
Starting with the $\Delta x^-$ integral and using 
(\ref{casebspacelike}), (\ref{casebwidth}), (\ref{denomDx+-}), we have 
analogous to (\ref{lDx+B}),
\be
\frac{l}{2} =\int_0^{r_*}dr\frac{(r/r_*)^{d-1}}{f(r,r_*)}
\frac{B-Qr^d}{\sqrt{B}\Big(1-\frac{Qr_*^d\left(1-(r/r_*)^{3d-2}\right)}
{2Bf^2}\Big)^{1/2}}\ .
\ee
As above, expanding to $O(Q)$ gives
\be
\begin{aligned}
\frac{l}{2} = \int_0^{r_*}dr \frac{(r/r_*)^{d-1}}{f(r,r_*)}
&+\frac{\delB}{2}\int_0^{r_*}dr \frac{(r/r_*)^{d-1}}{f(r,r_*)}\\
&+Qr_*^d \int_0^{r_*}dr \frac{(r/r_*)^{d-1}(1-(r/r_*)^{3d-2})}{4f^3(r,r_*)}-Q\int_0^{r_*}
dr \frac{r^d (r/r_*)^{d-1}}{f(r,r_*)}\ .
\end{aligned}
\ee
Then as above, the change in turning point is given by
\be\label{deltar*x-}
\begin{aligned}
-\delta r_* \eta  = \frac{r_* \delB}{2}\int_0^1 dx 
\frac{x^{d-1}}{\sqrt{1-x^{2(d-1)}}}
&+Qr_*^{d+1}\int_0^1 dx \frac{x^{d-1}(1-x^{3d-2})}{4(1-x^{2(d-1)})^{3/2}}\\
&-Qr_*^{d+1}\int_0^1 dx \frac{x^{2d-1}}{\sqrt{1-x^{2(d-1)}}}\ .
\end{aligned}
\ee
For this spacelike subsystem, the above (\ref{deltar*x-}) should 
be identical to (\ref{deltar*x+}). Using (\ref{pureadswidth}), 
this gives
\be
\delB = \alpha Qr_*^d\ ,\quad \mbox{with}\qquad  
\alpha = \frac{1}{\eta}\int_0^1dx \frac{x^{2d-1}}{\sqrt{1-x^{2(d-1)}}}
= \frac{\Gamma(\frac{1}{2d-2})\Gamma(\frac{1}{d-1})}{2(d-1)^2\Gamma(\frac{3}{2}+\frac{1}{d-1})\Gamma(\frac{d}{2d-2})}\ .
\ee
Using the above, we get
\be
\delta r_* = \beta Qr_*^{d+1}\ ,\qquad \mbox{with}\qquad
\beta = \frac{1}{4(d-1)} - 
\frac{2^{\frac{1}{d-1}}}{8(d-1)^3\sqrt{\pi}}\frac{\Gamma(\frac{1}
{2d-2})^2}{\Gamma(\frac{3}{2}+\frac{1}{d-1})}\ .
\ee
It can be checked that $\beta<0$\ ($\beta\ra 0^-$ for large $d$): 
thus $\delta r_*$ is negative.

We can do a similar perturbation for finding the $O(Q)$ change in the 
entanglement entropy $S_0$ for pure $AdS$ given by (\ref{pureadsee}). 
In the present $AdS_{d+1}$ plane wave case with the strip orthogonal to 
the flux, the entanglement entropy is (\ref{SEadspw}), \ie\
\be
4G_{d+1} S = 2V_{d-2}R^{d-1}\int_\epsilon^{r_*}\frac{dr}{r^{d-1}}
\frac{AB}{r^{d-1}\sqrt{\frac{A^2B^2}{r^{2(d-1)}}+Qr^d-2B}}\ .
\ee
From the turning point equation, we know that\ 
$AB=r_*^{d-1}\sqrt{2B-Qr_*^d}$. With $f(r,r_*)$ as defined before, the 
EE can be recast as
\be
4G_{d+1} S = 2V_{d-2}R^{d-1}\int_\epsilon^{r_*}\frac{dr}{r^{d-1}}
\frac{\left(1-\frac{Qr_*^d}{2B}\right)^{1/2}}{f(r,r_*)
\left[1-\frac{Qr_*^d(1-(r/r_*)^{3d-2})}{2Bf^2}\right]^{1/2}}\ .
\ee
Now with $B=1+\alpha Qr_*^d$, we see that the perturbation in EE is 
independent of $\delB$ to $O(Q)$, since $B$ appears above only as 
${Q\over B}$~. 
Expanding $S$ to $O(Q)$, we obtain
\bea
4G_{d+1} S &=& 2V_{d-2}R^{d-1}\int_0^{r_*} \frac{dr}{r^{d-1}}\frac{1}{f(r,r_*)}
\left[1-\frac{Qr_*^d}{4}+\frac{Qr_*^d}{4f^2}\left(1-(r/r_*)^{3d-2}\right)
\right]\nonumber\\
&=& 4G_{d+1} S_0 + 2V_{d-2}R^{d-1}Qr_*^2\int_0^1 dx\left[ \frac{1-x^{3d-2}}
{4x^{d-1}(1-x^{2(d-1)})^{3/2}}-\frac{1}{4x^{d-1}(1-x^{2(d-1)})^{1/2}}\right]
\nonumber\\
&=& 4G_{d+1} S_0\ +\ 2V_{d-2}R^{d-1}Qr_*^2 {\cal M}_{EE}\ ,
\eea
with
\be
{\cal M}_{EE} = \frac{\sqrt{\pi}}{4(d-1)^2}\left[\frac{\Gamma(\frac{1}{d-1})}
{\Gamma(\frac{d+1}{2d-2})}
-(d-1)\frac{\Gamma(\frac{d}{2d-2})}{\Gamma(\frac{1}{2d-2})}\right]\ .
\ee
It can be checked that ${\cal M}_{EE}>0$ for $d>1$. Thus the change in 
entanglement entropy is positive, as before. To $O(Q)$, we have 
$r_*\sim l$, so that as before, 
\be
\Delta S = {R^{d-1}\over 2G_{d+1}} {{\cal M}_{EE}\over 4\eta^2}\ V_{d-2} Ql^2\ 
=\ {R^{d-1}\over 2G_{d+1}} {{\cal M}_{EE}\over 4\eta^2}\ 
{V_{d-2}\over l^{d-2}} (Ql^d) \ ,
\ee
so that as in (\ref{deltaMI1}) previously, the mutual information 
decreases as
\be
I[A,B] = I_0[A,B]\ -\ 2 {R^{d-1}\over G_{d+1}} {{\cal M}_{EE}\over 8\eta^2}\ 
V_{d-2} Q l^2\ \left(1+{x\over l}\right)^2\ ,
\ee
in this perturbative regime with $Ql^d\ll 1$. 
It should not be surprising that no hint of the phase transition is 
visible in this perturbative regime. For subsystem size well below 
the characteristic length scale set by the energy density, \ie\ 
$l\ll Q^{-1/d}$, we only expect small corrections to the ground state 
entanglement and mutual information structure. The phase transition 
on the other hand corresponds to strips much wider than the 
characteristic length scale. In that regime, the two integrals for 
$\Delta x^\pm$ scale rather differently so that the spacelike 
subsystem requirement cannot be met: this leads to the absence of a 
connected surface and is the reflection of a phase transition.
The corresponding entanglement 
saturation occurs since the degrees of freedom responsible for 
entanglement do not straddle the boundary for long if their size 
$\sim O(Q^{-1/d})$ is much smaller than the subsystem width, since 
they enter the strip and leave.

\subsection{A more complete phase diagram and some numerical analysis}

\begin{figure}[h] 
\hspace{1pc} \includegraphics[width=18pc]{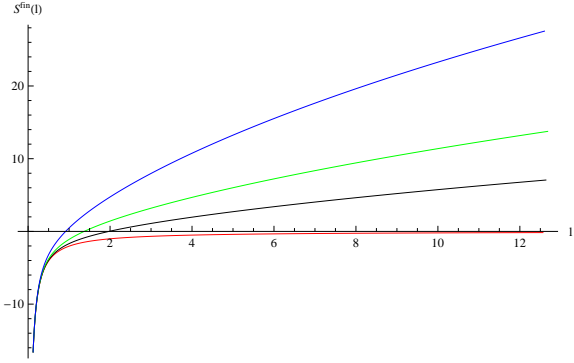} \hspace{2pc}
\begin{minipage}[b]{20pc}
\caption{{\label{figeeAdS4pw}\footnotesize {Plots of the finite parts of 
\newline entanglement entropy for the $AdS_4$ ground \newline
state (red) and the $AdS_4$ plane wave (the \newline 
black, green and blue curves correspond to \newline 
the values $Q=1, 3, 10$ respectively).  
\newline
}}}
\end{minipage}
\end{figure}

\begin{figure}[h] 
\hspace{1pc} \includegraphics[width=23pc]{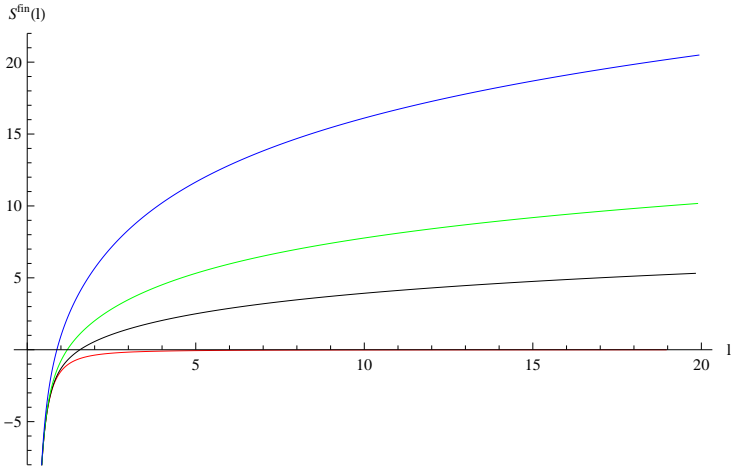} \hspace{2pc}
\begin{minipage}[b]{15pc}
\caption{{\label{figeeAdS5pw}\footnotesize {Plots of the finite 
\newline parts of entanglement entropy \newline 
for the $AdS_5$ ground state (red)  \newline and the $AdS_5$ plane wave 
(the \newline black, green and blue curves \newline 
correspond to the values \newline $Q=1, 3, 10$ respectively).
\newline \newline 
}}}
\end{minipage}
\end{figure}

In the previous subsections, we have studied entanglement entropy and
mutual information for large and small $Ql^d, Qx^d$. It is interesting
to study the interpolation between these, including the regime where
$Ql^d, Qx^d$ are $O(1)$. Towards this, we perform a numerical study of
the entanglement entropy integrals and thence mutual information
(using Mathematica). The plots in Figure~\ref{figeeAdS4pw} 
and Figure~\ref{figeeAdS5pw} show the finite cutoff-independent part
of entanglement entropy (black, green and blue curves) for the $AdS_4$ 
and $AdS_5$ plane waves, setting $Q=1, 3, 10$ respectively, in the 
case of the strip along the energy flux: the red curves are those for 
pure $AdS_4$ and $AdS_5$.  In the numerics, the area integrals have 
been regulated using a small UV cutoff regulator and subtracting off
the area law divergence term, we obtain the finite part.
For small $l$, we see that the $AdS$ plane wave (black, green, blue) 
curves lie ``above'' the pure $AdS$ (red) curves, which means the finite 
entanglement is larger than for the ground state.  This is of course 
consistent with the previous analytic studies in the perturbative and 
large $Ql^d$ regimes but the plots show that this is also true for all
$Ql^d$. Furthermore, the curves for larger $Q$ values lie ``above'' 
those for smaller $Q$ values, which is intuitively reasonable, implying 
that the finite entanglement increases with increasing energy density 
$Q$. The plot regions for large $l$ are in reasonable agreement
with fitted curves for $\sqrt{l}$ and $\log l$\ (the fits improve with
increasing accuracy, number of data points etc as expected with
numerics). 

\begin{figure}[h] 
\hspace{1pc} \includegraphics[width=18pc]{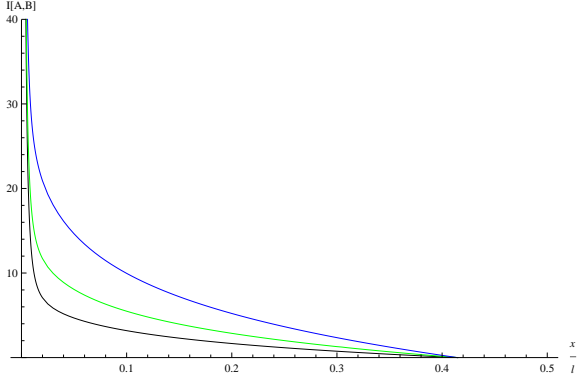} \hspace{2pc}
\begin{minipage}[b]{20pc}
\caption{{\label{figeeMIAdS5pw}\footnotesize {Plot of the mutual 
information \newline vs ${x\over l}$ with fixed width $l$ for the 
\newline $AdS_5$ plane wave. \newline \newline
}}}
\end{minipage}
\end{figure}

\begin{figure}[h] 
\hspace{1pc} \includegraphics[width=18pc]{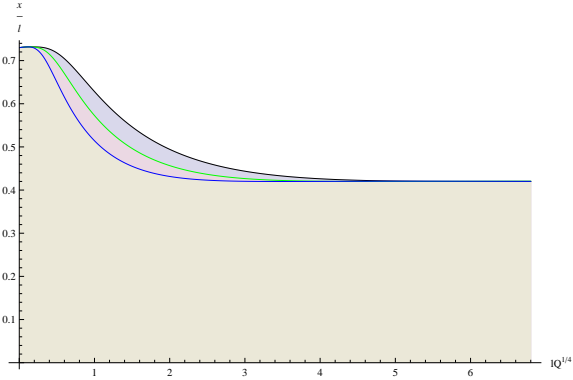} \hspace{2pc}
\begin{minipage}[b]{20pc}
\caption{{\label{figeeMIparamSpAdS5pw}\footnotesize {Plot of the 
${x\over l}$ vs $lQ^{1/d}$ parameter \newline 
space with nonzero mutual information \newline 
for the $AdS_5$ plane wave. \newline \newline
}}}
\end{minipage}
\end{figure}

Likewise, Figure~\ref{figeeMIAdS5pw} shows the plot of mutual
information vs the separation $x$ for the $AdS_5$ plane wave with both
strip subsystems along the flux (with fixed widths $l$ taken as
$l=50$). The small $x$ region shows a growth reflecting the divergence
when the subsystems approach to collide (which is similar to the
divergence for pure $AdS_5$). The mutual information vanishes at the
critical value ${x_c\over l}=0.41$. We have also checked that the
corresponding plot for pure $AdS_5$ behaves as expected, with the
critical value ${x_c\over l}\simeq 0.732$.
Figure~\ref{figeeMIparamSpAdS5pw} shows the ${x\over l}$ vs $lQ^{1/d}$ 
parameter space (shaded regions) with nonzero mutual information for 
the $AdS_5$ plane wave with both strip subsystems along the flux. We 
vary the width $l$ and find the critical value $x_c$ holding $Q$ 
fixed: the three curves are for $Q=1,3,10$ as before. We 
see that the critical value ${x_c\over l}$ interpolates from about 
$0.732$ ($lQ^{1/d}\ll 1$, approximately $AdS_5$ behaviour) to $0.41$ 
for the $AdS_5$ plane wave. We see that the mutual information 
parameter space remains nonzero for large $lQ^{1/d}$, unlike the finite 
temperature case \cite{Fischler:2012uv} where the curve has 
finite domain (with $x_c=0$ for large $lT$).
We have seen previously that in the wide strip regime $Ql^d\gg 1$, 
the mutual information disentangling transition location is independent 
of the energy density $Q$: this is reflected in 
Figure~\ref{figeeMIparamSpAdS5pw} by the fact that the black, green, 
blue curves all flatten out for large $l$, signalling that the 
critical value ${x_c\over l}$ is independent of the precise curve 
and corresponding $Q$ value. However we note that in the intermediate 
$Ql^d\sim O(1)$ regime, the mutual information disentangling 
transition location ${x_c\over l}$ certainly depends on the $Q$ value, 
the different curves being distinct. Thus it is only in the 
$Ql^d\gg 1$ regime that the mutual information disentangling transition 
becomes effectively independent of the energy flux $Q$.

There are similar plots for the $AdS_4$ plane wave, which we have 
not shown.

Our discussion so far and the corresponding plots have been for 
strips parallel to the energy flux. For the strips orthogonal to the 
flux, entanglement shows a phase transition, corroborated in the 
corresponding plot (shown in \cite{Narayan:2012ks}).
Plotting mutual information appears more intricate with more technical 
challenges in general. For wide strips $l\gtrsim Q^{-1/d}$, the strip 
entanglements saturate: crude plots show the strips disentangling at 
critical ${x_c\over l}$ values varying as $Q$ varies, with all 
${x_c\over l}$ less than those for the strip along the flux (\eg\ 
${x_c\over l}\sim 0.11$ with $Q=1$, $AdS_4$ plane wave). It would be 
interesting to study this more completely.

\section{Discussion}

We have studied entanglement entropy and mutual information in
$AdS_{d+1}$ plane waves dual to CFT excited states with energy-momentum 
density $T_{++}=Q$, building on \cite{Narayan:2012hk,Narayan:2012ks}, 
focussing on $d=3,4$ for two strips of width $l$ and separation $x$, 
parallel and orthogonal to the flux.

For the strips parallel to the flux, mutual information exhibits a 
disentangling transition at a critical separation ${x_c\over l}$ less 
than that for the ground state. For wide strips $Ql^d\gg 1$, we see that 
the subsystems disentangle only when they are sufficiently wide apart 
in comparison with the width: the critical separation ${x_c\over l}$ 
is independent of the characteristic energy scale $Q^{-1/d}$ in this 
regime. This is quite distinct from the finite temperature 
case \cite{Fischler:2012uv} where \eg\ the linear extensive growth of 
entanglement in the corresponding regime $lT\gg 1$ implies the 
subsystems disentangle for any finite separation $x$ independent of 
$l$. For the strips orthogonal to the flux, entanglement 
entropy shows a phase transition for $l\gg Q^{-1/d}$ \cite{Narayan:2012ks}: 
in this case, entanglement is saturated and so mutual information also 
vanishes.
In the perturbative regime $Ql^d\ll 1$ for the strips both parallel
and orthogonal to the flux, we have seen that the change in
entanglement entropy is $\Delta S\sim +V_{d-2} Q l^2$ with the
analysis similar to ``entanglement thermodynamics''. Here the mutual
information always decreases. Thus the disentangling transition 
in this regime again occurs for separations smaller than those for the
ground state. In this perturbative regime, the critical separation
${x_c\over l}$ certainly depends on $Q$ and $l$.
The numerical study shows the critical ${x_c\over l}$ has nontrivial 
dependence on $Q$ in intermediate regimes as well. As one approaches 
the wide strip regime $Ql^d\gg 1$, the mutual information curves 
approach each other and flatten out, signalling independence with $Q$.

Overall this suggests that the energy density disorders the system, so
that the subsystems disentangle faster relative to the ground
state. The thermal state is disordered, since in the regime with
linear (extensive) entropy, the subsystems are disentangled or
uncorrelated for any nonzero separation $x$. The $AdS$ plane wave
states are in some sense ``partially ordered'': the disentangling
transition location occurs at critical values ${x_c\over l}$ smaller
than those for the ground state for the strip along the energy flux, 
but the critical value remains nonzero even for wide strips 
$Ql^d\gg 1$. Perhaps this ``semi-disordering'' is also true for more 
general excited states that are ``in-between'' the ground and 
thermal states.

The $AdS_5$ plane wave gives rise to a hyperscaling violating
spacetime exhibiting logarithmic violation of entanglement entropy,
suggesting that perhaps these are indications of Fermi surfaces
\cite{Ogawa:2011bz,Huijse:2011ef}. In the regime where the strip
widths and separation are large relative to the energy scale
$Q^{-1/4}$, the logarithmic scaling of entanglement implies a
corresponding scaling of mutual information, similar to the 
corresponding behaviour for Fermi surfaces.  This regime is
of course just one part of the full phase diagram thinking of these as
simply excited states in $AdS/CFT$, as we have seen. It would be
interesting to explore these further.

\vspace{3mm}
\noindent {\small {\bf Acknowledgments:} It is a pleasure to thank 
M. Rangamani, T. Takayanagi and S. Trivedi for comments on a draft, 
and M. Headrick for comments on a previous version of the paper. 
KN thanks J. Maldacena for an early conversation pertaining to a 
parton-like description for $AdS$ plane waves, while on a visit to 
the IAS, Princeton, and the IAS for hospitality over that period.}

\end{document}